\documentclass[epj]{webofc}
\usepackage[varg]{txfonts}   
\woctitle{Physics at the Magnetospheric Boundary}
\begin{document}
\title{The magnetosphere of the close accreting PMS binary V4046 Sgr}
%
%

\author{S. G. Gregory\inst{1}\fnsep\thanks{\email{sg64@st-andrews.ac.uk}} \and
V. R. Holzwarth\inst{2} \and
J.-F. Donati\inst{3} \and
G. A. J. Hussain\inst{4} \and
T. Montmerle\inst{5} \and
E.~Alecian\inst{6} \and
S. H. P. Alencar\inst{7} \and
C. Argiroffi\inst{8} \and
M. Audard\inst{9} \and
J. Bouvier\inst{10} \and
F. Damiani\inst{8} \and
M. G{\"u}del\inst{11} \and
D.~P.~Huenemoerder\inst{12} \and
J. H. Kastner\inst{13} \and
A. Maggio\inst{8} \and
G. G. Sacco\inst{14} \and
G. A. Wade\inst{15}
}

\institute{School of Physics \& Astronomy, University of St Andrews, St Andrews, KY16 9SS, U.K. 
\and
Freytagstr. 7, D-79114 Freiburg i.Br., Germany
\and
UPS-Toulouse/CNRS-INSU, Inst. de Recherche en Astro. et Plan{\'e}tologie, F-31400 Toulouse, France \and 
ESO, Karl-Schwarzschild-Str. 2, D-85748 Garching, Germany \and
Institut d'Astrophysique de Paris, 98bis bd Arago, FR-75014 Paris, France \and
Obs. de Paris, LESIA, 5, place Jules Janssen, F-92195 Meudon Principal Cedex, France \and 
Dept. de F{\`i}sica - UFMG, Av. Ant{\^o}nio Carlos, 6627, 30270-901 Belo Horizonte, MG, Brazil \and
INAF-Osservatorio Astronomico di Palermo, Piazza del Parlamento 1, I-90134 Palermo, Italy \and
ISDC Data Center for Astrophysics, Univ. of Geneva, CH-1290 Versoix, Switzerland \and
UJF-Grenoble 1/CNRS-INSU, IPAG, UMR 5274, F-38041, Grenoble, France \and
Dept. of Astronomy, University of Vienna, Trkenschanzstrasse 17, A-1180 Vienna, Austria \and 
MIT, Kavli Inst. for Astrophysics \& Space Research, Cambridge, MA 02139, U.S.A. \and
CIS, Rochester Inst. of Technology, 54 Lomb Memorial Drive, Rochester, NY 14623, U.S.A. \and 
INAF-Arcetri Astrophysical Observatory, Largo Enrico Fermi 5, I - 50125 Florence, Italy \and
Dept. of Physics, Royal Military College of Canada, Kingston, K7K 7B4, Canada
}

\abstract{
V4046 Sagittarii AB is a close short-period classical T Tauri binary. It is a circularised and synchronised 
system accreting from a circumbinary disk. In 2009 it was observed as part of a coordinated program involving 
near-simultaneous spectropolarimetric observations with ESPaDOnS at the Canada-France-Hawai'i 
Telescope and high-resolution X-ray observations with {\it XMM-Newton}. Magnetic maps of each star 
were derived from Zeeman-Doppler imaging. After briefly highlighting the most significant observational 
findings, we present a preliminary 3D model of the binary magnetosphere constructed from the magnetic 
maps using a newly developed binary magnetic field extrapolation code. The large-scale fields (the dipole components) of both 
stars are highly tilted with respect to their rotation axes, and their magnetic fields are linked.
}
\maketitle
\section{Introduction - V4046 Sgr AB}
\label{intro}
V4046 Sgr is a nearby ($\sim$73$\,{\rm pc}$) pre-main sequence (PMS) binary\footnote{V4046 Sgr may 
be a quadruple system \cite{kas11}.  \cite{tor06} first speculated that the distant GSC 07396-00759 
(projected separation $\sim$12,350$\,{\rm au}$) was a companion to V4046 Sgr AB. Our detailed study 
of this weakly bound object revealed that it is likely a spectroscopic weak-line T Tauri binary system, and 
was christened V4046 Sgr C[D] by \cite{kas11}.}, and a candidate member of the $\beta$ Pictoris Moving 
Group \cite{tor06}. It is a double lined spectroscopic binary that is still accreting from a large circumbinary 
disk \cite{ros12,ros13}, despite being of an age ($\sim$13$\,{\rm Myr}$) at which the disks of most other PMS stars have dispersed (e.g. \cite{mam09}). At this age, both stars of the system have developed 
substantial radiative cores within their interiors \cite{don11,gre12}. The stellar parameters are listed in 
table~\ref{binpara}. Below we present the main observational findings from a multi-wavelength observing 
program which targeted V4046 Sgr in 2009, and give brief details of a newly developed binary magnetic 
field extrapolation code that we are using to model the binary magnetosphere.  

\subsection{A near-simultaneous multi-wavelength observing program}
\label{obs}
Near-simultaneous optical and X-ray monitoring of V4046 Sgr was carried out in September 2009, 
covering $\sim$2.5 and $\sim$2.2 orbital cycles respectively.  The main goal of the program was to 
characterise the 3D magnetic and accretion flow geometry of the system, as well as the properties of the 
X-ray emitting plasma.  Time-resolved optical/X-ray spectra are essential for this purpose, as previously
demonstrated for the accreting PMS star V2129 Oph \cite{donv2129,arg11}. 
 
{\it Optical observations}: spectropolarimetric observations were obtained with ESPaDOnS at the Canada-France-Hawai'i 
Telescope from 3-9 September 2009 \cite{don11}.  Small rotationally modulated circular polarisation 
signals were detected in the photospheric absorption lines, but (unusually for accreting PMS stars) no 
signal was measured in the accretion-related emission lines suggesting that accretion hotspots may be 
distributed across the stellar surface (consistent with their complex magnetic fields, see below).

\begin{table}[t]
\centering
\caption{Stellar parameters for V4046 Sgr. $M_{\rm core}$, the mass of the radiative core, $M_\ast$ \& 
age are derived from the models of \cite{sie00}. As a synchronised system the stellar rotation periods are 
the same as the binary orbital period.}
\label{binpara}
\begin{tabular}{llllllll}
\hline
Star & $T_{\rm eff}$ & $L_\ast/{\rm L}_\odot$ & $M_\ast/{\rm M}_\odot$ & age (Myr) & $M_{\rm core}/M_\ast$ & $P_{\rm rot}$ (d) & Ref. [$T_{\rm eff}$,$L_\ast$,$P_{\rm rot}]$ \\\hline
V4046 Sgr A & 4370 & 0.407 & 0.91 & 13.0 & 0.47 & 2.42 & \cite{ste04,don11,qua00} \\                       
V4046 Sgr B & 4100 & 0.269 & 0.87 & 13.0 & 0.40 & 2.42 & \cite{ste04,don11,qua00} \\\hline
\end{tabular}
\end{table}

\begin{figure}
\centering
  \includegraphics[height=0.305\textwidth]{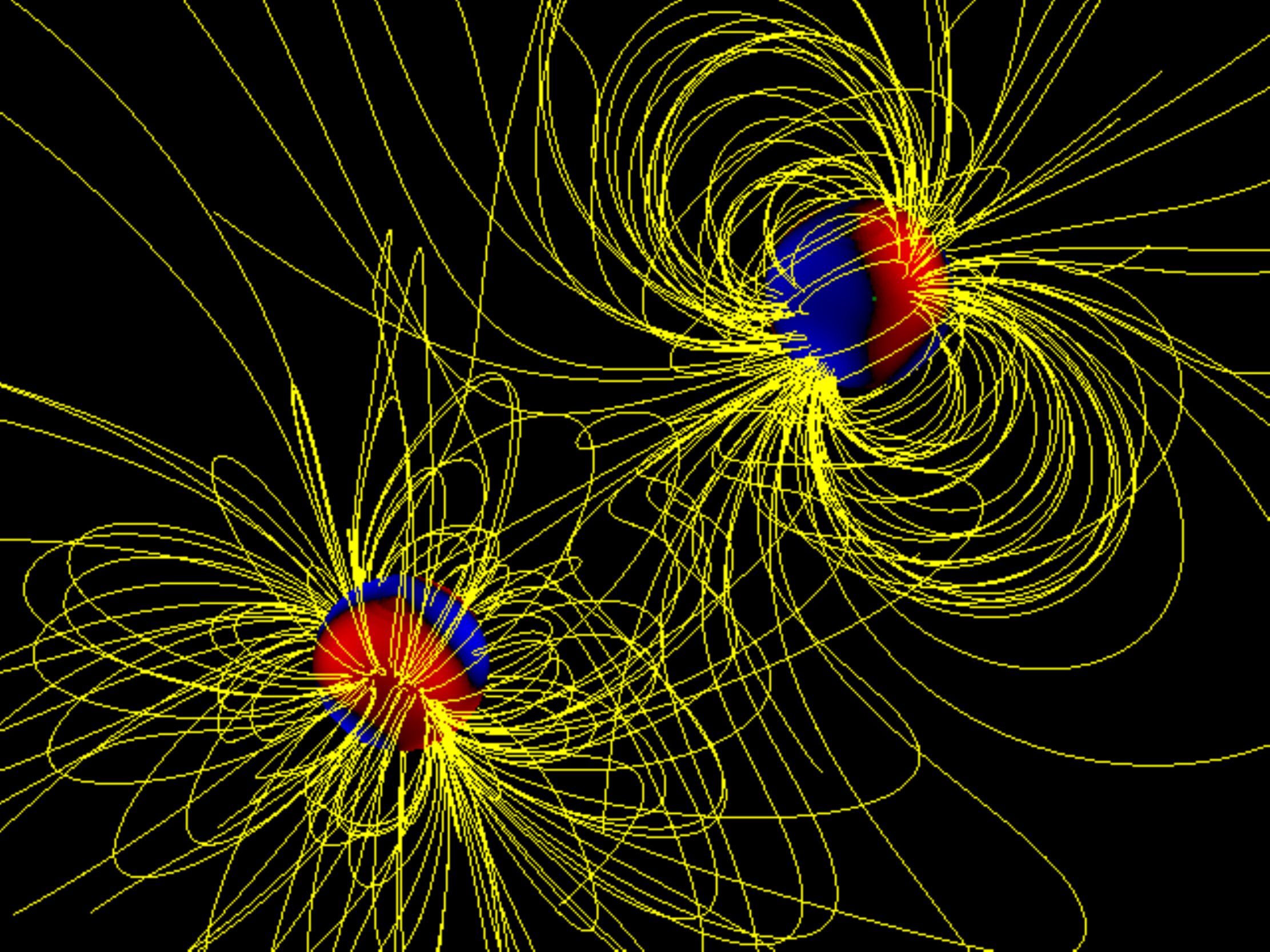}
    \includegraphics[height=0.305\textwidth]{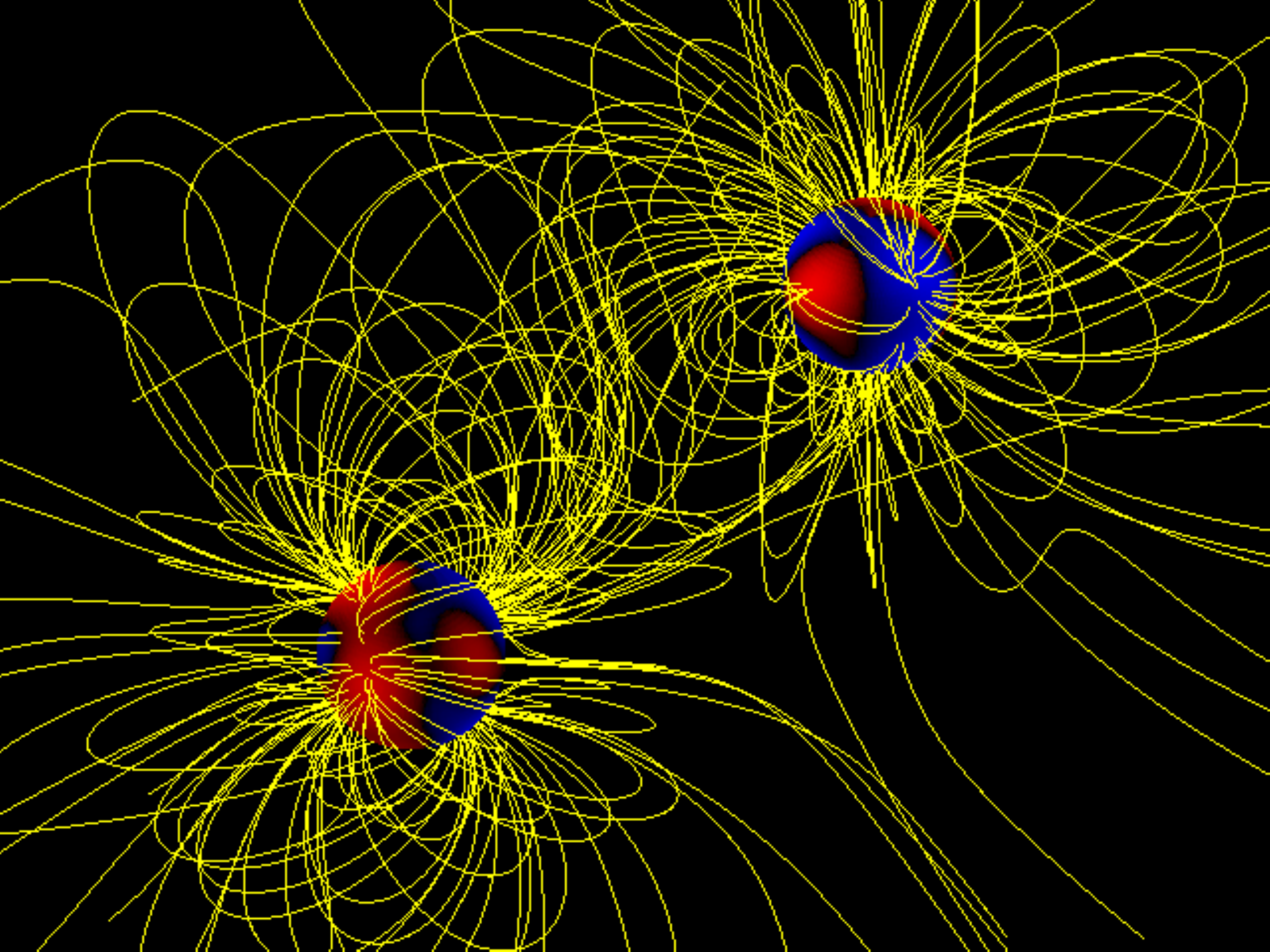}
\caption{A binary potential field extrapolation from the magnetic maps of V4046 Sgr at two different 
phases.  Only the large-scale field lines are shown for clarity. The magnetic fields of the two stars are 
linked with field lines connecting the dayside of the primary star (shown in the background/foreground of 
the left/right image) to the nightside of the secondary. The binary separation is 
$\sim$8.8$\,{\rm R}_\odot$ \cite{don11} and the system is believed to be synchronised and circularised 
(eccentricity $\le0.01$; \cite{ste04}).}
\label{extrap}  
\end{figure}

A tomographic imaging code was applied to the circularly polarised spectra to produce magnetic maps of 
each star \cite{don11}; the first such maps to be obtained for a close accreting PMS binary.  Both stars 
host highly complex large-scale magnetic fields \cite{don11}, as has been found for other PMS stars with 
similar interiors consisting of large radiative cores and outer convective envelopes \cite{gre12}. The 
magnetic fields are multipolar with strong toroidal components, while the poloidal components are mostly 
non-axisymmetric.  In particular, the dipole components are weak, of polar strength 
$\sim$100$\,{\rm G}$ \& $\sim$80$\,{\rm G}$ on the primary/secondary, and highly tilted relative to the rotation axes ($\sim$60$^{\circ}$ \& $\sim$90$^\circ$) with a phase difference of $\sim$0.7 between the planes of the 
tilts \cite{don11,gre12}.  
        
{\it Xray observations}: the {\it XMM-Newton} observations of V4046 Sgr, consisting of three $\sim$123$\,{\rm ks}$ observing 
segments separated by $\sim$50$\,{\rm ks}$ gaps, were obtained between 15 \& 20 September 2009 
\cite{arg12}.  Strong X-ray emission from cool plasma of temperatures $3-4\,{\rm MK}$ and number 
densities $10^{11}-10^{12}\,{\rm cm}^{-3}$ was detected (as it was previously by \cite{gun06}). This
plasma, which is cooler and denser than is typical of coronal plasma, is associated with accretion 
shocks, where gas channelled along magnetic field lines slams into the stellar surface at supersonic 
speeds. 

The emission lines in the X-ray grating spectra produced by the high density plasma displayed periodic 
flux variations with a period of $1.22\pm0.01\,{\rm d}$, half of the binary orbital period of $2.42\,{\rm d}$ 
\cite{arg12}.  This is the first time that rotationally modulated accretion shock X-ray emission has been 
detected from a PMS star.  At first glance this appears inconsistent with the optical observations, as it 
suggests discrete accretion hotspots.  However, the accretion column/spot geometry may indeed be 
complex as X-rays from accretion shocks can only be detected at certain viewing angles when hotspots 
are not obscured by the dense accreting gas \cite{arg11}.

\section{The magnetosphere of V4046 Sgr AB extrapolated from magnetic maps}
\label{field}
A new binary magnetic field extrapolation code has been developed, allowing
for the construction of 3D models of the joint magnetosphere of binary stars
from observationally derived magnetic maps~\cite{hol13}. 
The models are generated in the framework of the potential field-source
surface approximation, in which the magnetosphere is taken to be current-free
and the effect of magnetised winds is mimicked through magnetic fields being
purely radial on the source surface.

The magnetic field, $\vec{B}= -\nabla\Psi$, in the current-free region is
determined by the solution of Laplace's equation, $\nabla^2 \Psi= 0$, subject
to boundary conditions given on the source surface and on the surface of the
two stellar components. The starting point for the binary field extrapolation is the \emph{ansatz},
\begin{equation}
\Psi (\vec{r})= 
\sum_{\ell m} 
\left(
\alpha_\ell^m a_\ell^m
+
\beta_\ell^m b_\ell^m
\right)
+
\sum_{\ell m} 
\left(
\gamma_\ell^m g_\ell^m
+
\delta_\ell^m d_\ell^m
\right)
+
\sum_{\ell m}
\left(
\epsilon_\ell^m e_\ell^m
+
\zeta_\ell^m z_\ell^m
\right).
\label{ansatz}
\end{equation}
The three sums on the RHS describe the contributions of the source surface,
primary surface and secondary surface to the total potential $\Psi$, where
$a_\ell^m (\vec{r}_0), b_\ell^m (\vec{r}_0), g_\ell^m (\vec{r}_1), d_\ell^m (\vec{r}_1),
e_\ell^m (\vec{r}_2)$, and $z_\ell^m (\vec{r}_2)$ are solid spherical harmonics with
$\vec{r}_0, \vec{r}_1$, and $\vec{r}_2$ being vectors connecting the centre of the spheres
representing the source surface, the primary star, and the secondary star, respectively, to
the point $\vec{r}$. 
By applying a least-squares fit of observed magnetic surface maps and source
surface conditions, we derive the equation 
\begin{equation}
\mathbf{A} \left( \begin{array}{c} \vec{\alpha} \\ \vec{\beta} \\ \vec{\gamma} \\ \vec{\delta} \\ 
\vec{\epsilon} \\ \vec{\zeta} \end{array} \right) =
\mathbf{B} \left( \begin{array}{c} \vec{\chi} \\ \vec{\sigma} \end{array} \right)
+
\mathbf{C} \left( \begin{array}{c} \vec{\rho} \\ \vec{\varsigma} \end{array} \right)
\label{maeq}
\end{equation}
where the matrices $\mathbf{A}, \mathbf{B}$, and $\mathbf{C}$ contain sums of
products of solid spherical harmonics and their gradients, which depend only
on the geometrical properties of the binary/source surface system.
The vectors $\vec{\chi}, \vec{\sigma},\vec{\rho}$ and $\vec{\varsigma}$ on the
RHS of equation~(\ref{maeq}) are the known expansion coefficients of the magnetic
maps of the primary and secondary star.
The system of equations~(\ref{maeq}) is solved for the expansion coefficients
$\vec{\alpha},\vec{\beta},\vec{\gamma},\vec{\delta},\vec{\epsilon}$, and
$\vec{\zeta}$ using a singular value decomposition (SVD) algorithm.
Full details of the binary magnetic field extrapolation method will be 
given in a forthcoming paper \cite{hol13}.

A field extrapolation from the magnetic maps of V4046 Sgr \cite{don11} is
shown in figure~\ref{extrap}.
The large-scale magnetic field of each star is highly tilted, and there are
``S-shaped'' field lines connecting through the interior region of the binary
from the dayside of the primary to the nightside of the secondary
\cite{gre13}. With such a field configuration it is unlikely that local
circumstellar disks, distinct from the global circumbinary disk, can form
around each star; as was suggested for V4046 Sgr based on hydrodynamic
simulations \cite{dev11}.  This will be addressed in a future paper, and
presented alongside detailed calculations of the accretion spot/flow geometry
of the system \cite{gre13}.

\begin{acknowledgement}
SGG acknowledges support from the Science \& Technology Facilities Council (STFC) via an Ernest 
Rutherford Fellowship [ST/J003255/1]. GAW is supported by a Discovery Grant from the Natural Science \& Engineering Research Council of Canada (NSERC).
\end{acknowledgement}

%
%
%

\end{document}